\documentstyle[multicol,aps,prl,psfig,epsf]{revtex}

\begin{document}
\twocolumn[\hsize\textwidth\columnwidth\hsize\csname
 @twocolumnfalse\endcsname

%\begin{document}
%\draft
\title{Elastic interaction and superstructures in manganites and
other Jahn-Teller systems}

\author{D.~I.~Khomskii$^{a}$ and K.~I.~Kugel$^{b}$}
\address{$^{a}$Laboratory of Solid State Physics, Groningen University,
\\
 Nijenborgh 4, 9747 AG Groningen, The Netherlands\\
$^{b}$Institute for Theoretical and Applied Electrodynamics, Russian Academy
of Sciences,
\\
Izhorskaya Str. 13/19, 127412 Moscow, Russia}
\date{\today}
\maketitle

\begin{abstract}
The role of elastic interactions between Jahn-Teller ions in formation of
 various orbital- and charge-ordered structures in manganites and related
 compounds is analyzed. It is shown that such interactions alone are often
 sufficient to reproduce the structures observed in different regions of the
 phase diagram. A special attention is focused on stripe structures at high
 doping levels.
\end{abstract}
\smallskip
\vskip2pc
]

%\pacs{64.75.+g, 61.50.Ah, 71.28.+d}
%\tighten
%\bigskip

\section{Introduction}

As is well established now on many examples, an orbital degeneracy
 and Jahn-Teller effect play very important role in various
 transition-metal and rare-earth compounds
 \cite{Goodenough,KaplVekh,KugKhUFN}. In concentrated systems,
 it leads typically to lowering of lattice symmetry, often taking
 the form of a structural phase transition; it is associated with
 (or caused by) the orbital ordering. Correspondingly, also
 magnetic properties are largely determined by the type of occupied
 orbitals according to the Goodenough-Kanamori-Anderson rules,
 see e.g.\ \cite{Goodenough,KhSawats,KhSpinEl}.

Manganites La$_{1-x}M_{x}$MnO$_{3}$ ($M$ = Ca, Sr, Ba) with the
 colossal magnetoresistance (CMR) form a very popular nowadays system,
 for which the orbital effects often play very important role.
 There may be other rare earths (Pr, Nd, Sm) instead of La in these
 series; there exist also layered analogues  of three-dimensional
 perovskite manganites, often having similar properties. It
 is established that there exists an orbital ordering of some
 kind in almost all  parts of the phase diagram of these systems.
 Thus, it exists in  La$_{1-x}$Ca$_{x}$MnO$_{3}$ at $x=0$;
 probably in the charge ordered insulating phases at small
 $x$ $(x\sim 0.1-0.2)$; at the checkerboard-like charge ordered
 phase at $x=0.5$ (so called CE phase); in the overdoped regimes
 $0.5\leq x\leq 1$, where charge and orbital stripes
 \cite{Radaelli,Raveau} or bistripes \cite{ChenCheong} were observed.
 The only exception may be the ``optimally doped'' ferromagnetic metallic
 phase $0.25\leq x\leq 0.5$, where the very phenomenon of CMR is
 taking place: no apparent orbital ordering of usual type is observed
 there at low temperatures \cite{Egami} although the possibility
 of an orbital ordering of a new type (with  ``complex orbitals'')
 was suggested even for this phase \cite{vdBrinkKh}.
 In the present paper, we will not discuss these more exotic
 possibilities and will limit ourselves to the conventional
 types of orbital ordering with real orbitals.

There exist in principle two different mechanisms of orbital ordering,
 or cooperative Jahn-Teller effect. First, it is the usual interaction
 of degenerate electrons with the crystal lattice
 \cite{KaplVekh,Englman,EnglHalp}, which is usually considered
 as a source of the Jahn-Teller effect. In concentrated
 systems, this interaction provides the coupling of orbital occupation
 at different sites and can lead to an orbital ordering simultaneously
 with the corresponding lattice distortion. The second mechanism is the electronic (exchange) one \cite{KugKhJETP,KugKhUFN}, which can give rise
 to  both the spin and orbital orderings, and which was quite successful in
 explaining the magnetic and orbital structures in a number of materials,
 notably cuprates \cite{KugKhUFN}.Most of the previous theoretical
 considerations were confined to undoped systems containing integer number
of electrons at the transition metal sites: $d^{9}$ for Cu$^{2+}$,
$d^{4}$ for Mn$^{3+}$. An important class of systems is represented
 by the doped oxides like La$_{1-x}$Ca$_{x}$MnO$_{3}$, where, as
 mentioned above, similar phenomena of orbital ordering are also often
 observed. One should generalize the models used for undoped systems
 to such cases, in particular to be able to explain different
 superstructures observed in these systems, e.g., stripes or paired
 stripes (bistripes).

Such treatment was recently initiated in \cite{KugKhEPL}, where we
considered the  formation of superstructures, including stripes,
 due to elastic interaction, when we dope our system, i.e.\ substitute
 ions of one valence (and one size) by an ``impurity'' with the
 different valence and different atomic volume. It was shown there
 that due to a specific nature of these interactions (long-range
 character, $\sim 1/R^{3}$, and, most importantly, different sign --
repulsion or attraction -- in different directions) the structures
 of different kinds -- 1D stripes, 2D ``sheets'' -- can be naturally
 formed in these cases.

When considering the case of Jahn-Teller (JT) systems, one should
 modify this treatment by taking into account the anisotropic nature
 of corresponding electronic states, i.e.\ anisotropic electronic
 charge distribution(having  quadrupolar character) and, correspondingly,
 the shape of a Jahn-Teller ion in a respective orbital state.
 Thus, in contrast to the treatment of  \cite{KugKhEPL}, we have
 to consider not the interaction of spherical  impurities
 (``sphere-in-the-hole'' model) but rather that of anisotropic
 impurities (``ellipsoid-in-the-hole''). This generalization was
 shortly mentioned at the end of \cite{KugKhEPL}; in the present
 paper, we consider this situation in detail and discuss the
 applications to the superstructures observed in manganites in
 different regions of the phase diagram. As we shall
 see, in some cases such interactions alone are sufficient to
 reproduce the observed structures. In other situations, some extra
 factors have to be invoked, but in any case one can say that
 the elastic interactions considered in this paper are
 definitely very important in stabilizing the observed
 orbital and, consequently, magnetic structures.

\section{Strain mechanism of superstructure formation}

First, we shortly summarize the results concerning the ordering of
 ``spherical'' impurities due to elastic forces \cite{KugKhEPL}.
 When we put into the host matrix an ion of different size, e.g.\ by
 substituting an ion with the atomic volume $v_{0}$ by an impurity
 with the volume $v\neq v_{0}$, this creates a strain field, which
 decays rather slowly, as $1/R^{3}$ \cite{Eshelby}. Another such
 impurity interacts with this strain, which therefore leads to a certain
 coupling between impurities. If both the host and impurity ions are
 spherical, this is what is called ``sphere-in-the-hole'' model. It is known
 that the strain-induced interaction vanishes for the isotropic media
 \cite{Eshelby} (except of the infinite-range interaction caused by
 the ``mirror forces'' related to the sample surface). However, the real
 crystals are always anisotropic. This gives rise to an interaction,
 which has the following form for weakly anisotropic cubic crystals
 \cite{KugKhEPL,Eshelby}:
 \begin{equation}
V({\bf r},{\bf r}^{\prime })=
 -
\frac{CQ_{1}Q_{2}d\left( n_{x}^{4}+n_{y}^{4}+n_{z}^{4}-\frac{3}{5}\right) }{
R^{3}}
  \eqnum{1}
 \end{equation}
 Here $Q_{i}=v_{i}-v_{0}$ are the ``strengths'' of the impurities,
 $R=|{\bf r}-{\bf r}^{\prime }|$ is the distance between them,
 $n_{x}$, $n_{y}$, and $n_{z}$ are the direction cosines
 of vector $\bf R$.
 The most important parameter entering (1) is
\begin{equation}
d=c_{11}-c_{12}-2c_{44}  \eqnum{2}
\end{equation}
 where $c_{ij}$ are elastic moduli of the crystal. This parameter
 carries the information about anisotropy: the case $d=0$
 corresponds to an isotropic medium.

 The most significant feature of the interaction (1) is
 that it is attractive in certain directions independent of
 the impurity type ($v>v_{0}$ or
 $v<v_{0}$) and of the sign of the coefficient $d$. Thus for $d>0$ the
 interaction is attractive along cubic axes [100], [010], and [001] and
 repulsive along face and body diagonals of the cubic cell, [110] and
 [111], and vice versa for $d<0$. As argued in \cite{KugKhEPL}, this
 quite naturally leads to the formation of superstructures (1D stripes,
 2D sheets) in insulating systems: the second, third, etc.\ impurities
 ``migrate'' toward the first one along certain directions, e.g.\
 [001] for $d>0$, and finally form vertical (diagonal for $d<0$) stripes
 along this direction.
 Important is also that here this motion of impurities is provided by
 the electron hopping and does not require real  diffusion of atoms;
 thus, e.g., holes moving in nickelates transform Ni$^{2+}$ ions
 into Ni$^{3+}$ ``impurities'', which can form stripes due
 to this mechanism.

 When considering such processes in systems with Jahn-Teller ions,
 e.g.\ in manganites, one has to generalize this treatment to the case
 of anisotropic impurities -- instead of ``sphere-in-the-hole'' we have
 to consider ``ellipsoid-in-the-hole''. In this case, there would be
 two factors determining the interaction between such centers: one is
 the elastic anisotropy of the host lattice, and another, specific
 for the JT case, is the dependence of the interaction on
 a relative orientation
 of the corresponding orbitals. In general, one has to consider both
 factors, but in the weakly anisotropic crystals it is often the
 second factor, which plays the dominant role, whereas the first
 one may differ from system to system.

\section{The model and relevant interactions}

 We consider below the situation with doubly degenerate $e_g$ orbitals,
 with one electron on them (e.g., ions Mn$^{3+}$ ($t_{2g}^{3}e_{g}^{1}$),
 low-spin Ni$^{3+}$ ($t_{2g}^{6}e_{g}^{1}$)) or with one $e_g$ hole,
 Cu$^{2+}$ ($t_{2g}^{6}e_{g}^{3}$). As the basis orbitals one can take
 $d_{z^{2}}= \left| 3z^{2}-r^{2} \right \rangle $ and
 $d_{x^{2}-y^{2}}= \left| x^{2}-y^{2} \right \rangle $. The electron
 occupation of these orbitals corresponds to a quadrupolar distribution
 of electron density; the elongated electron ellipsoid for the electron
 $z^{2}$ orbitals, and the flattened (compressed) ellipsoid for
$x^{2}-y^{2}$
 orbitals. One may expect also the corresponding displacement of the
 ligands (e. g., the distortion of O$_{6}$ octahedra): local elongation
 for $z^{2}$ electron and $x^{2}-y^{2}$ hole orbitals, and local compression
 in the opposite case. These local distortions will lead to a strain field
 in a crystal, which, as discussed above,
will provide the mechanism of coupling between such ions. The latter can
 finally lead to the formation of one or another type of superstructures.
 This is the main mechanism considered in this paper.

 There are, in principle, several aspects in the electron-lattice
 interactions. The strain-induced interaction mentioned above is the
 interaction via long-wave\-length pho\-nons. This elastic interaction depends
 on the type of ``impurity'' and also on the anisotropy of the crystal.
 For anisotropic impurities, like JT centers considered in this paper,
 this interaction depends not only on the relative direction between
 sites in a crystal, but also on the relative local distortions on each
 site. As a result, this interaction becomes very complicated
 \cite{Eremin,Fishman}.
 In addition to long-range interactions mediated by the strain field in a
 crystal, there exists also the interaction mediated by the short-range, or
 optical, vibrations. The detailed form of these interactions was studied in
 \cite{EnglHalp}, and below we shall take these interactions into account.
 It turns out that the general structure of both these contributions is
 rather similar: the interactions between the nearest-neighbor sites along
 crystal axes [100], [010], and [001], which can be obtained from
 \cite{Eremin} and \cite{EnglHalp}, are exactly the same, and the
 interactions of next nearest neighbors along the diagonals [110],
 [011], etc.\ have the same sign, although somewhat different ratios.
 Thus, there are two contributions,
 which we consider below, the contribution due to the long-wavelength
 phonons (strain interaction), and that due to the nearest neighbor
 coupling via short-wavelength or optical phonons.

 Generally speaking, one has to include also the electronic terms describing
 the exchange contributions to orbital ordering \cite{KugKhUFN} and also an
 electron hopping, which can lead to delocalization of electrons. We
 consider below only the insulating states with localized electrons;
 therefore, we ignore electron hopping here, although in principle the
 tendency to delocalization and formation of metallic states can compete
 with the states considered below.

 There is one assumption that we make below. In general, an arbitrary
 combination of the basic orbitals $z^{2}$ and $x^{2}-y^{2}$ of the type
$$\left| \theta \right \rangle =\cos \theta /2 \left|
z^{2}\right\rangle + \sin \theta /2 \left| x^{2}-y^{2} \right
\rangle$$
is allowed, corresponding both to local elongation and
 contraction of  O$_{6}$ octahedra or to any combination
 of these deformations (which
 would give both prolate and oblate ellipsoids or even biaxial ones).
 Experimentally it is well established, however, that only the local
 elongations are realized in practice: out of hundreds known
 Jahn-Teller compounds
 there are virtually none with compressed octahedra (for localized
 electrons) \cite{KugKhUFN}. There are also physical reasons for that,
 related to the lattice anharmonicity  and higher-order interactions
 \cite{Kanamori,KhvdBr-anh}. In accordance to that, we will consider
 below only the situation with locally elongated ``impurities'',
 but the axes of local elongations in cubic crystals may be directed
 along $x$, $y$, or $z$ axes ([100], [010], or [001]), i.e.\ angles
 $\theta $ may be $0$, $\pm 2\pi /3$ for the one-electron orbitals
 as in Mn$^{3+}$, and $\theta = \pi $, $\pm \pi /3$ for the one-hole
 orbitals at Cu$^{2+}$. Thus, we consider
 the case of elongated impurities (ellipsoids) interacting with one
 another via short-wavelength and long-wavelength phonons. The
 corresponding situations are illustrated in Figs.\ 1a and 1b for
 the neighbors along cubic axes $x$, $y$, and $z$, and in
 Figs.\ 1c and 1d -- for diagonal neighbors. It is qualitatively
 clear from these
 figures that the situation of Fig.\ 1a ($z^{2}$ orbitals at
 nearest-neighbor pair along $z$ axis) would correspond to the
 repulsion (intermediate oxygen O1 feels ``conflicting'' forces),
 whereas that of Fig.\ 1b would give an attraction.
 This is indeed confirmed by the actual calculations \cite{EnglHalp},
 where the constants of these two and of two other possible
 situations were determined, see 1)~--~4) in Table 1 (this is
actually the interaction via short-wavelength, or optical phonons).

The situation is somewhat more complicated for diagonal pairs
 like those in Figs.\ 1c and 1d. Here, it is the noncentral forces,
 which determine
 the resulting interactions. Still it is qualitatively clear that the
 interaction of oxygens O2 and O3 in Figs.\ 1c and 1d would lead to an
 effective repulsion in the first case and to an attraction in the second
 one (note that actually, in contrast to our figure, oxygen ions are the
 biggest ones, so that O2 and O3 ``hard spheres'' touch one another,
 which explain our conclusion). The coupling constants corresponding
 to Figs.\ 1c  and 1d and for two other ``diagonal'' configurations
 (cases 5)~--~8) in Table~1) we also take from \cite{EnglHalp}.

 \setbox0=\hbox{\cite{EnglHalp}}

%\begin{center}
\begin{table}%[tbp]
\hspace{-0.0cm}
\epsfxsize=8.5cm
\centerline{\epsffile{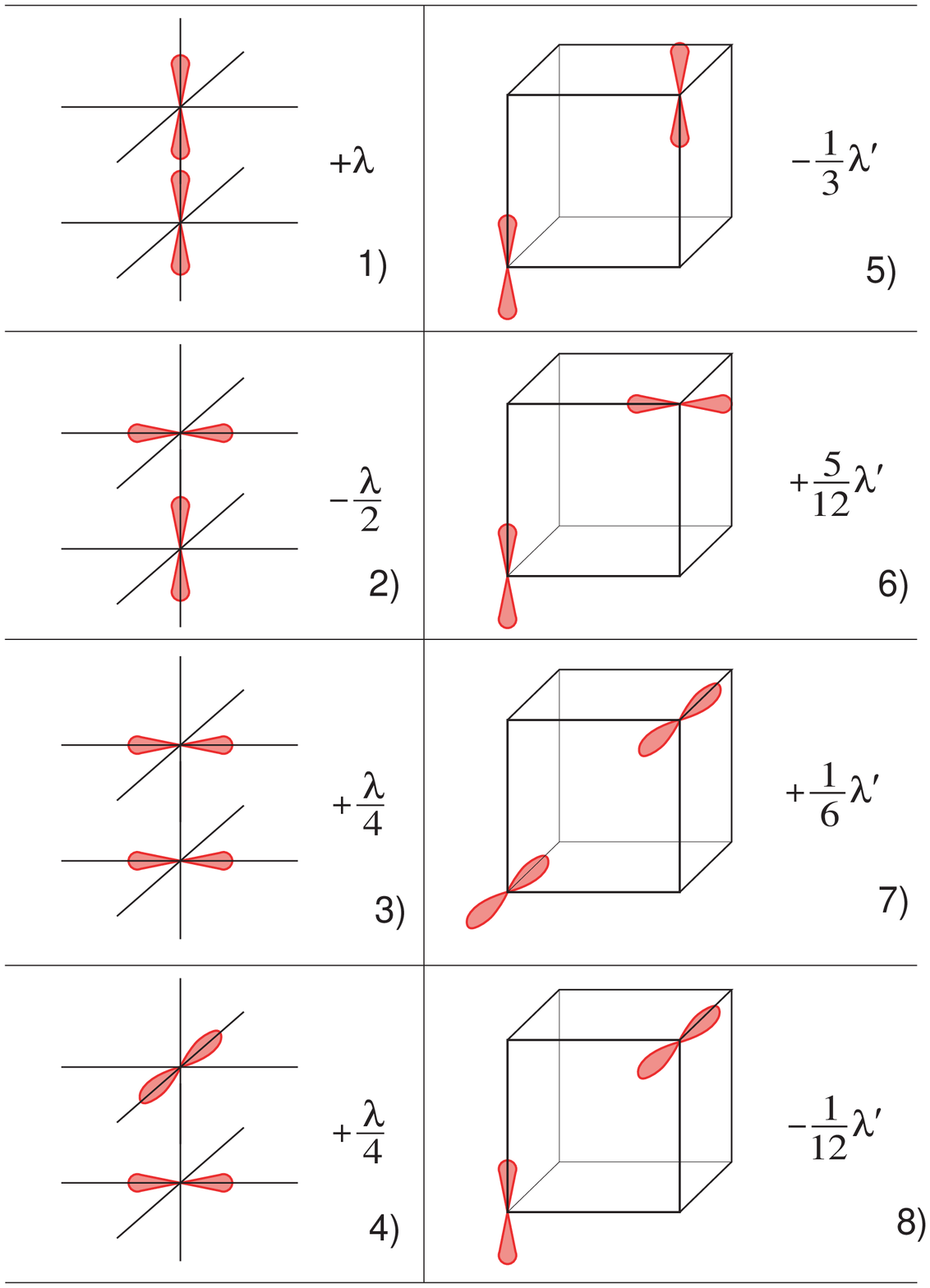}}
 \vskip1\baselineskip plus1\baselineskip minus.3\baselineskip
\caption{The interaction between different orbitals (local distortions)
 for nearest and next nearest neighbors. These interactions
 are expressed through two independent constants, $\lambda$ and
 $\lambda^{\prime }$, characterizing central (cases 1) -- 4))
 and noncentral (cases 5) -- 8)) forces. The form and ratios of
 these interactions are taken from\box0; the long-range
 elastic interactions (3) give essentially the same expressions,
 see text. Here the electron orbitals are shown, in which case the
 local distortions have the same form (elongation along the electron
 cloud). For the one-hole case one should draw the orthogonal hole
 orbitals (e.g.\ in the case 1) -- the hole orbitals $x^2-y^2$).}
\label{Table}
\end{table}

Thus, the resulting
 nearest-neighbour interactions 1) - 8) are expressed through two
 parameters, $\lambda $ and $\lambda ^{\prime }$, corresponding to the
 central and noncentral forces, respectively. Note also that these
 interactions take into account both the lattice structure (lattice
 anisotropy) and the anisotropic character of orbital occupation and
 of the on-site distortions.

%%%%%%%%%%%%%% Fig 1 %%%%%%
\begin{figure}[tbp]
\epsfxsize= .95\hsize  \vskip 1.0\baselineskip
\centerline{\epsffile{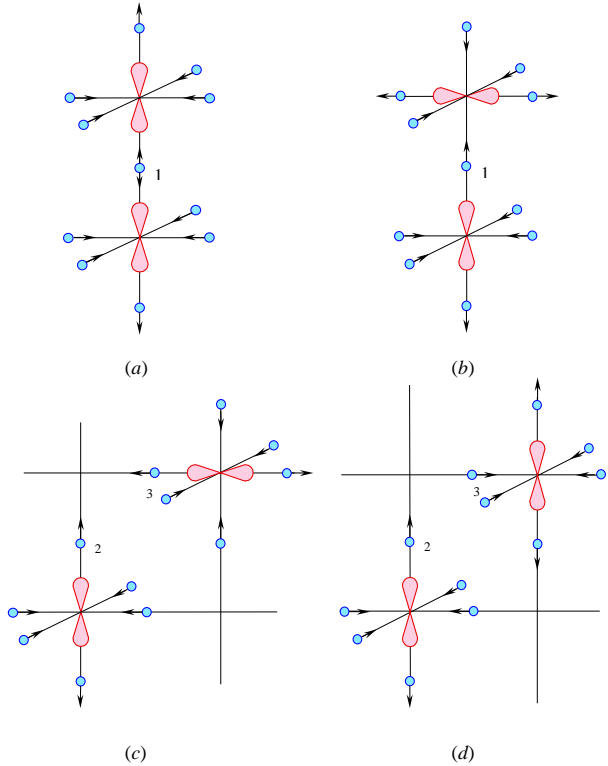}}
\caption{Interactions between two Jahn-Teller ions with different
 occupied orbitals: (a) and (b) -- pair along $z$ axis; (c) and (d)
 -- pair along the face diagonal of the perovskite unit cell.
 Displacements of oxygen ions are indicated by arrows.}
\label{pairs}
\end{figure}

The second contribution, due to interaction via the long-wavelength
 phonons for the case of impurities in the isotropic medium (i.e.,
 for zero value of parameter $d$ defined by (2)) can be obtained from
 the general expression presented in \cite{Eremin}; it has the form

\begin{eqnarray}
V&=&\frac{(c_{11}+c_{44})}{8\pi(c_{11}+2c_{44})R^3}\Bigl\{5\sigma_{zz}^{(1)}
 \sigma_{zz}^{(2)}+ \nonumber\\
 &&+2\Bigl(\sigma_{xx}^{(1)}\sigma_{xx}^{(2)}+\sigma_{yy}^{(1)}
 \sigma_{yy}^{(2)}\Bigr)\Bigr\}+\nonumber\\
 &&+\frac{1}{4\pi R^3}
\Bigl(2\sigma_{zz}^{(1)}\sigma_{zz}^{(2)}-\sigma_{xx}^{(1)}\sigma_{xx}^{(2 )
}
 -\sigma_{yy}^{(1)}\sigma_{yy}^{(2)}\Bigr)  \eqnum{3}
 \end{eqnarray}
 where $c_{11}$ and $c_{44}$ are elastic moduli, and
 $\sigma_{\alpha\alpha}$ is a stress tensor such that e.g.\
for the center with the occupied orbital $3z^2-r^2$ we have
$\sigma_{zz}=1$, $\sigma_{xx}=\sigma_{yy}=-1/2$ (and
 corresponding expressions for $z \to x, y$).

 The calculations show that the signs and the ratios of the interaction
 constants for the nearest neighbours (cases 1) -- 4) in Table 1)
 obtained from (3), exactly coincide with those via short-wavelength
 phonons presented in Table 1. For the "diagonal" interactions (cases 4)
 -- 8) in the Table 1) the signs of the interactions obtained from (3)
 are the same as those given in the Table 1, although the ratios of
 these constants differ somewhat. This is probably due to the
 assumption of the elastic isotropy made in (3) (constant $d$ (2)
 is put $=0$), which is apparently less satisfactory for the
 noncentral forces.

 For $d\neq 0$, the corresponding expression becomes much more
 complicated (see \cite{Eremin} for details). Often, for analysis
 of elastic interactions of impurities, further simplifications
 are introduced, namely the longitudinal and transverse sound velocities
 are taken to be equal. This leads to a rather simple explicit
 expression for the interaction between Jahn-Teller ions with
 different occupied orbitals, situated at arbitrary positions with
 respect to crystal axes \cite{Fishman}. For some problems (e.g.\ for
 the analysis of ESR spectra), such simplifications work reasonably
 well. However, in our case such approximation is definitely
 insufficient, leading to unphysical degeneracy in energies of
 different orbital configurations. At the same time, formula (3), as
 it was already mentioned, qualitatively correctly reproduces the
 main features of pair interactions even for nearest and diagonal
 neighbours in the cubic crystal.

 Below, in calculating the energies of different ordered states,
 we would take into account only two types of interactions between
 the nearest neighbouring JT ions: those along cubic axes $x$,
 $y$, $z$, and along face diagonals. For this purpose, we will
 use the interaction constants presented in Table 1
 \cite{EnglHalp}, because they take into account both the anisotropy
 of JT ions and the elastic anisotropy of the lattice.

 As we can see from Table 1, the interactions are determined by two
 independent constants: the first one, $\lambda $, corresponds to
 the directions along the cube axes (cases 1) -- 4)) and is determined
 by central forces between ions, and another one, $\lambda ^{\prime }$
 -- along diagonals of the elementary plaquettes -- is caused by the
 noncentral forces \cite{EnglHalp}. For the elastic interactions,
 these constants are determined by different elastic moduli of the
 crystal: by the bulk and shear moduli, respectively.

 In principle, one should include also longer-range interactions, but
 their due account would require rather extensive numerical calculations,
 which we postpone for future publications.
 Still, we will show below that even using only the nearest neighbor
 interactions, we can successfully describe many superstructures
 observed e.g. in manganites at different doping levels, and can
 analyze such questions as the relative stability of single {\em vs}
 paired stripes, compare the energy of different types of orbital domain
 walls, etc. One may hope that the longer-range interactions, which
 are still much weaker than those taken into account, would not modify
 our main conclusions.

 \setbox0=\hbox{{\bf LaMnO$_3$}}

\section{\box0, undoped layered manganites and other
 similar materials}

 First, we consider the simplest case of undoped oxides, which may be
 relevant for such systems as undoped perovskite LaMnO$_{3}$ or
 KCuF$_{3}$, layered materials like K$_2$CuF$_4$, etc.
These systems contain ions Mn$^{3+}$ ($t_{2g}^{3}e_{g}^{1}$)
 or Cu$^{2+}$ ($t_{2g}^{6}e_{g}^{3}$) at each lattice site,
 forming a simple cubic lattice for LaMnO$_{3}$ and
 square perovskite-like lattice in layered systems.

 {\bf IVa}. First, we consider the 2D square lattice relevant
 for layered systems. One immediately sees that if the
 $\left| 3z^{2}-r^{2} \right \rangle $ and $\left| x^{2}-y^{2} \right
\rangle $
 orbitals are exactly degenerate, then with the interactions
 presented in Table 1 one would get the structure shown in
 Fig.~2. This structure is stabilized first of all by the attraction
 ($-\lambda /2$) of orthogonal orbitals (deformations) $x^{2}$ and
 $y^{2}$ along $x$ and $y$ directions. The diagonal coupling is that
 of parallel orbitals (case 5) in Table 1), i.e.\ it is also
 attractive ($-\lambda ^{\prime }/3$) and gives extra stabilization
 of this configuration. Its energy (per site) is

\begin{equation}
E=2(-\lambda /2)+2(-\lambda ^{\prime }/3)=-\lambda -2\lambda ^{\prime }/3.
\eqnum{4}
\end{equation}

 One can easily see that all the other possible orderings of these
 local distortions (locally elongated octahedra) give higher energy.
 The structure described above coincides with the ordering observed in
 the layered cuprate K$_{2}$CuF$_{4}$ \cite{Yto} (which was predicted
 earlier on different grounds \cite{KhKugSSC}). The only difference is
 that for Cu$^{2+}$ such a packing of local distortions corresponds not
 to an alternation of electron orbitals $x^{2}$ and $y^{2}$, but to the
 corresponding pattern formed by the hole orbitals ($y^{2}-z^{2}$ and
 $x^{2}-z^{2}$). The same ordering is also observed in layered
 chromites Rb$_{2}$CrCl$_{4}$, see \cite{KugKhUFN,RbCuCl3}.

 Due to a layered structure of corresponding systems there may exist in
 them an initial splitting of the $z^{2}$ and $(x^{2}-y^{2})$-orbitals corresponding to local elongation along $c$-direction, and
 if this splitting is large enough, it can give the ferro-orbital
 ordering such as the $x^{2}-y^{2}$ ordering in La$_{2}$CuO$_{4}$.
 For layered manganites (single-layer La$_{1-x}$Sr$_{1+x}$MnO$_{4}$,
 bilayer La$_{2-2x}$Sr$_{1+2x}$Mn$_{2}$O$_{7}$), we would expect in
 this case the $z^{2}$ occupation at each cite. This is apparently
 the situation in the undoped ($x$ = 0) layered manganites
 \cite{z^2order}.

%%%%%%%%%%%*****************FIG.2*******************%
 \begin{figure}[tbp]
\epsfxsize= .95\hsize  \vskip 1.0\baselineskip
\centerline{\epsffile{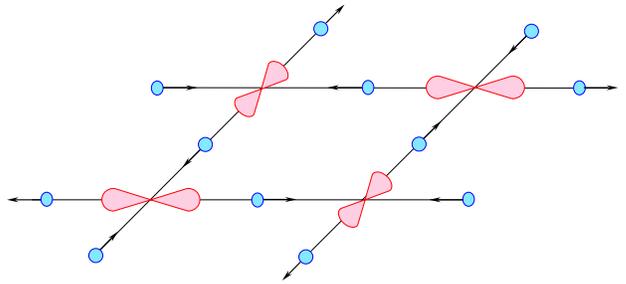}}
\caption{Possible packing of local distortions and orbital structure
(occupied electron orbitals) for undoped layered systems. Oxygen
 shifts are shown by arrows} \label{xy_plane}
 \end{figure}
%%%%%%%%%%%%

 {\bf IVb}.
 Let us now consider the case of a cubic perovskite LaMnO$_{3}$
 (we ignore at first the tilting of MnO$_{6}$ octahedra leading
 to the orthorhombicity). One immediately sees that the same
 factors, which acted in layered systems, would stabilize the
 superstructure of Fig.~2 in the basal planes of LaMnO$_{3}$.
 The complication arises, when one considers the ordering pattern
 in the third ($z$ axis) direction. One can have two types of
 arrangement of such planes. They can form either the in-phase
 structure (Fig.~3a), or the out-of-phase structure (Fig.~3b).
 These are usually referred to as the $d$-type and $a$-type
 structures, respectively. With the interactions of Table 1, their
 energies turn out to be the same:
\begin{eqnarray}
E_{3a}&=&2 \left( - \frac{\lambda}{2} \right) + \frac{\lambda}{4} +
 2 \left( - \frac{\lambda ^{\prime }}{3} \right) +
 4 \left( - \frac{\lambda ^{\prime }}{12} \right)\nonumber \\
&=&- \frac{3}{4} \lambda - \lambda ^{\prime } \eqnum{5}
\end{eqnarray}

\begin{eqnarray}
E_{3b}&=&2 \left( - \frac{\lambda}{2}\right) + \frac{\lambda}{4} +%
2 \left( - \frac{\lambda ^{\prime }}{3}\right) +%
2 \frac{\lambda^{\prime }}{6} +%
2 \left( - \frac{\lambda ^{\prime }}{3}\right)\nonumber \\
&=&- \frac{3}{4} \lambda - \lambda ^{\prime } \eqnum{6}
\end{eqnarray}

Experimentally, both types of ordering are observed in perovskites:
 ordering of the $d$-type (Fig.~3a) is found e.g.\ in
 LaVO$_{3}$, and the $a$-type structure -- in YTiO$_{3}$.
 In KCuF$_{3}$ both types of ordering may arise, and stacking faults
 are formed very easily. For LaMnO$_{3}$, only the ordering of
 the $d$-type (Fig.~3a) is observed. Apparently, there are some
 other factors, not  included on our model, which determine
 the ordering in third direction.

%%%%%%%%%%%*****************FIG.3******************
\begin{figure}[tbp]
\epsfxsize= .95\hsize  \vskip 1.0\baselineskip
\centerline{\epsffile{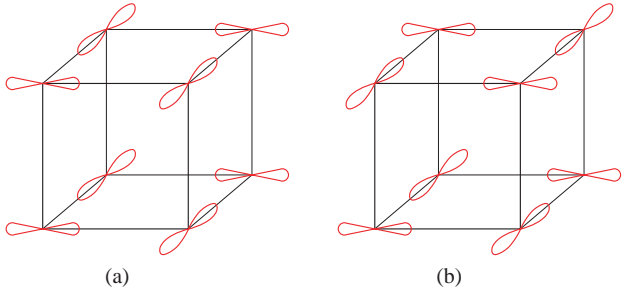}}
 \vskip1\baselineskip plus1\baselineskip
\caption{Two different kinds of orbital ordering in cubic
manganites: (a) in-phase and (b) out-of-phase.} \label{twotypes}
 \end{figure}

%%%%%%%%%%%%

One such factor may be an interplay of the  Jahn-Teller distortions
 with the GdFeO$_{3}$-type distortion caused by the tilting of the
 MeO$_{3}$ octahedra \cite{Mizokawa} -- one may show that for strong
 enough tilting this interplay stabilizes the $d$-type ordering
 of Fig.~3a. In any case, the main motive -- packing of the
 distortions shown in Fig.~2 -- is the first and the most important
 ingredient in the orbital superstructures observed in many
 perovskites with JT ions, and this directly follows from our
 mechanism.

Thus, we see that the electron-lattice coupling quite naturally gives
the correct lattice and orbital superstructure of undoped manganites
(see also \cite{EnglHalp,Novak})

\section{Low-doped manganites}
 There are several interesting systems, which show rather puzzling
 properties at low doping level. Thus, La$_{1-x}$Sr$_{x}$MnO$_{3}$
 has an insulating ferromagnetic state at about $x=1/8
 (\sim 0.1\leq x\leq 0.14$) \cite{JapLaSr(1/8),BuechLaSr(1/8)}.
 Such a state appearing in oxides is in general a rare occasion:
 most of the insulating systems are antiferromagnetic, and ferromagnetism
 usually coexists with (and is explained by) the metallicity (double
 exchange).  Probably, the only way to stabilize a ferromagnetic
 insulating state  is by forming a special orbital ordering
 \cite{KhSawats}, so as to give ferromagnetic exchange according
 to the aforementioned Goodenough-Kanamori-Anderson rules.
 The superstructure was indeed observed in  La$_{1-x}$Sr$_{x}$MnO$_{3}$
 in this region \cite{Yamada}, and it was interpreted in \cite{Yamada}
 as predominantly the charge ordering -- ordering of holes (Mn$^{4+}$
 ions) of more or less $bcc$ type in the Mn$^{3+}$ matrix. The remaining
 Mn$^{3+}$ ions, of course, should have certain orbital occupation in an
 insulator, which could in principle finally lead to ferromagnetism.

This problem was considered theoretically in \cite{Mizokawa1}, where
 it was shown that if one assumes the charge ordering pattern of
 Yamada {\em et al.} \cite{Yamada}, one would indeed obtain certain
 orbital ordering, which can be treated as an ordering of orbital polarons
 \cite{Kilian,Mizokawa2} leading to a ferromagnetic spin ordering.
 These orbital polarons are the objects, in which the central small-size
 Mn$^{4+}$ ion is surrounded by larger Mn$^{3+}$ ions, with the lobes of
 their orbitals directed toward the central Mn$^{4+}$. The elastic forces
 are definitely rather important in forming such objects and in their
 possible ordering (see also \cite{AhnMillis}). However, the Hartree-Fock
 calculations carried out in \cite{Mizokawa1} have shown that the better
 structure with the lower energy is the one, in which the $xy$ planes
 alternate as follows: one plane is a pure Mn$^{3+}$ plane with the
 superstructure of  Fig.~2; the next plane contains all the holes
 (Mn$^{4+}$ ions) forming ``vertical'' stripes (stripes along $x$
 or $y$ directions), see Fig.~4. Such a state turned out to be also
 ferromagnetic; this structure is consistent with the lattice symmetry
 observed in \cite{Yamada}.

 Although the numerical calculations of \cite{Mizokawa1} were carried
 out using different starting point, one can argue that it is the
 special stability of the orbital ordering of Fig.~2 that helps to
 stabilize (or is even a driving force of) such a superstructure
 in low-doped manganites, with segregation of pure undoped Mn$^{3+}$
 planes, all the holes being in other planes. If it is true, one can
 expect similar phenomena also in other systems, e.g.\ in low-doped
 Pr$_{1-x}$Ca$_{x}$MnO$_{3}$, where the ferromagnetic insulating phase
 is also known to exist at $0.1\leq x\leq 0.3$ \cite{Tomioka,Jirak}.
 It is possible in principle that the structure of this system, e.g.\
 close to $x=1/4$, is also formed by the ordered array of orbital polarons
 \cite{Mizokawa2}. But the better alternative may be again the
segregation of holes in every second plane, undoped planes being indeed
 like those in Fig.~2, and doped planes (which in this case would have
 the effective doping $x_{\text{eff}}=n_{\text{holes}}=0.5$) having
 either stripe-like structure or even the superstructure of the CE type,
 characteristic of $x$ = 0.5 (see below).

%%%%%%%%%%%*****************FIG.4 *******************%
\begin{figure}[tbp]
\epsfxsize= .95\hsize  \vskip 1.0\baselineskip
\centerline{\epsffile{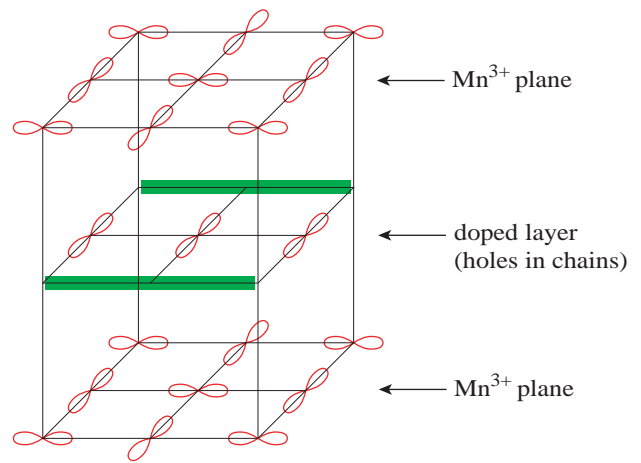}}
\vskip0.3\baselineskip plus1\baselineskip
\caption{A possible orbital ordering at $n = 1/8$.}
\label{n_0_25}
\end{figure}

%%%%%%%%%%%%

 The detailed type of charge
 and orbital ordering in Pr$_{1-x}$Ca$_{x}$MnO$_{3}$ in the vicinity
 of $x=1/4$ is not known yet, although the first measurements by the
 anomalous X-ray scattering \cite{Zimmermann} have shown that in this system
 there indeed exists a superstructure, which can be consistent with
 that in Fig.~4 and/or with the CE type superstructure of Fig.~5a.
 It would be very interesting to check this possibility experimentally
 in more detail.

\section{Half-doped systems: Charge and orbital ordering}

\noindent {\bf VIa}. {\em Ground state}

In most of half-doped manganites $R_{1-x}$Me$_{x}$MnO$_{3}$, $x=0.5$ ($R$ = La, Pr, \dots),
 there exist at low temperatures the charge and orbital ordering
 accompanied by the antiferromagnetic ordering of the CE type
 \cite{Wollan,Jirak}, the latter usually appearing at still lower
 temperatures. Nowadays, one often uses the term ``CE ordering'' also
 to denote the corresponding charge and orbital structures. This ordering
 is illustrated in Fig.~5a. It consists of the checkerboard arrangement
 of Mn$^{3+}$ and Mn$^{4+}$ ions with the corresponding orbital ordering
 at Mn$^{3+}$ sites. Of course, one should not take notations ``3+'' and
 ``4+'' too literally: actual degree of charge disproportionalization can
 be much less, e.g., $\sim (3.5\pm 0.2)$ \cite{vdBrKhalKh}. This is the
 structure in the basal plane; the stacking of these planes in the third
 direction is in phase, so that 3+ and 4+ sites, as well as the
 corresponding orbitals, are on top of one another.

%%%%%%%%%%%*****************FIG.5*******************%
\begin{figure}[tbp]
\epsfxsize= .91\hsize  \vskip 1.0\baselineskip
\vbox{\centerline{\epsffile{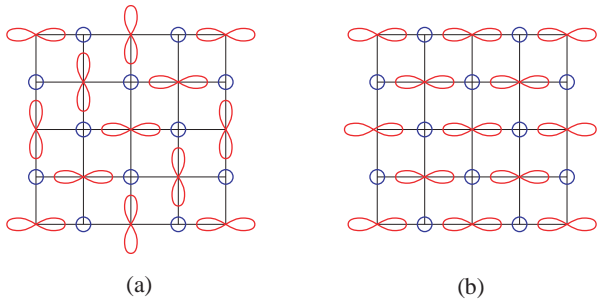}}
\vskip0.3\baselineskip plus1\baselineskip
\caption{Charge and orbital orderings in half-doped manganites: (a) actual
 ordering (the so called CE structure); (b) alternative orbital
 ordering with all occupied orbitals at Mn$^{3+}$ of the same type
 ($x^{2}$ orbitals). Empty circles denote Mn$^{4+}$ ions, the
 electron density distribution for the corresponding orbitals is shown
 for the Mn$^{3+}$ sites.}}
\label{orborCE}
\end{figure}

There are two questions related to this superstructure. The first one
 concerns the type of charge ordering. It is quite natural to expect the
 checkerboard arrangement of Mn$^{3+}$ and Mn$^{4+}$ ions, or of extra
 electrons: such ordering is actually analogous to a Wigner crystal,
 i.e.\ it is favored by the Coulomb forces. These considerations,
 however, are definitely not enough, because in this model one would
 expect charge (valence) alternation also in the third direction,
 which is not the case. The solution of this problem may again be
 connected with the elastic interactions: it can be shown
 \cite{KugKhEPL,Khach} that they typically favor the inclusion of one
 phase (e.g.\ Mn$^{4+}$) in another (Mn$^{3+}$) in the form of
 infinitely thin slabs -- sheets, oriented in a crystal in a
 particular way, so as to minimize total strain energy. These may
 be just Mn$^{4+}$ sheets observed in the CE structure.

 Another question concerns the type of orbitals occupied at Mn$^{3+}$
 sites. If we assume the checkerboard charge ordering, we may have
 different types of occupied orbitals, e.g., one can in principle get
 the ordering shown in Fig.~5b with all orbitals of the same type
 (here, $x^{2}$ orbitals) instead of alternation of $x^{2}$ and
 $y^{2}$ diagonal rows (``stripes'') of Fig.~5a.

 The type of orbital ordering in the checkerboard charge structure can
 be again analyzed using the interactions listed in Table 1. Here we also
 include the interaction of two types of nearest neighbors: the first
 neighbors in the Mn$^{3+}$ sublattice are here the diagonal interactions
 5) -- 8) of Table 1, and the second neighbors are those along $x$ and
 $y$ directions in Fig.~5. The only difference is that these
 interactions are now at twice the distance of the nearest neighbors
 Mn$^{3+}$ and Mn$^{4+}$ in these directions, so that according to
 the general nature of elastic interactions, which decay as
 $1/R^{3}$, these couplings with parameter $\lambda $ (cases 1) -- 4)
 in Table 1) should be now multiplied by factor $1/2^{3}=1/8$.

From Table 1 one sees that the diagonal interactions are attractive for
 the same orbitals, e.g., $x^{2}$ and $x^{2}$ (case 5) in Table 1), but
 repulsive for $x^{2}$ and $y^{2}$ (case 6)). From this point view,
 one should rather expect that the orbital structure in this case would
 correspond to Fig.~5b, and not to Fig.~5a. However, the interactions
 along $x$ and $y$ directions tend to stabilize the structure of
 Fig.~5a. One can easily calculate the energies of these two
 competing states (per site):
\begin{equation}
E_{5a} = \left( \frac{-\lambda ^{\prime }}{3}\right) +\left(%
 \frac{5\lambda ^{\prime }}{12}\right) +2\cdot%
 \frac{1}{8}\left( \frac{-\lambda }{2}\right) =%
 \frac{\lambda ^{\prime }}{12}-\frac{\lambda }{8}  \eqnum{7}
 \end{equation}
 Here, the first two terms are the diagonal interactions with the first
 neighbouring Mn$^{3+}$ ions and the last term is the interaction along
 $x$ and $y$ directions (at a distance of two lattice parameters).

 Similarly, the energy of the state presented in Fig.~5b is
\begin{equation}
E_{5b}=2\left( \frac{-\lambda ^{\prime }}{3}\right)%
 + \frac{\lambda }{8}+\frac{1}{8}\left( \frac{-\lambda }{4}\right)%
 = -\frac{2\lambda ^{\prime }}{3} + %
\frac{5\lambda }{32}  \eqnum{8}
\end{equation}
 Thus, if the central force interaction $\lambda $ is strong enough,
\begin{equation}
\lambda ^{\prime }/\lambda < 3/8 =0.375,  \eqnum{9}
\end{equation}
 the CE structure of Fig.~5a would be stable; for the opposite
 inequality one would get the structure of Fig.~5b. Since actually
 the CE structure is observed in experiment, one should conclude that
 inequality (9) is fulfilled in manganites (within the approximations
 made). One can also show that all other feasible types of orbital
 orderings have higher energy.

 One has to stress that the assumption of the checkerboard charge ordering
 is crucial for this conclusion; if we would not make this assumption and
 use only the JT-induced interactions of Table 1, one should rather get in
 the model under study the phase separation: all electrons (Mn$^{3+}$ ions)
 would form a dense cluster with the superstructure of Fig.~2. This
 would minimize the interaction energy of the JT-distorted sites. Thus
 one has to add some extra factors to this model, e.g., the Coulomb
 interaction preventing such phase separation. However, assuming the
 checkerboard charge ordering, we can explain the concomitant orbital
 ordering of Fig.~5a in our model. (Nevertheless, the possibility of phase separation  cannot be completely discarded even in this case
 -- and all the more so at $x\neq0.5$ \cite{Kagan}.)

% NEW SUBSECTION :
\noindent {\bf VIb}. {\em Orbital domains}

It is also of interest, and of practical importance, to study
 in our model the ``cost'' of creation of defects, notably orbital
 domain walls. As argued in \cite{orbdomains} and especially in
 \cite{Hill}, such domain walls are easily formed in CE states, e.g.\
 in
La$_{0.5}$Ca$_{0.5}$MnO$_{3}$ and in Pr$_{1-x}$Ca$_{x}$MnO$_{3}$.

 One can easily calculate in our model the energy of such defects.
 The simplest one is the local defect -- rotation of the orbital on
 one site, e.g.\ from $x^{2}$ to $y^{2}$. With our interaction
 constants it would cost us $\Delta E=9/16 \lambda$ (diagonal
 interactions remain the same
for the real CE structure corresponding to Fig.~5a).
 More interesting is the situation with the domain walls. There are four
 types of them, classified and drawn in Figs.\ 3b --~3e of \cite{Hill}
. Domains separated by domain walls 3b and 3c of \cite{Hill} are
 those with the same propagation vector, i.e.\ the same orientation of
 orbital stripes, and 3d and 3e are domain walls between two domains
 with the propagation vectors rotated by 90 degrees (orbital twins).
 Straightforward calculations show that the energies of domain walls
 (per unit length, i.e.\ per Mn$^{4+}$-site at the center of domain
 wall) are
\begin{equation}
\Delta E_{3b}=-\frac{3}{4}\lambda ^{\prime }+\frac{9}{32} \lambda
\eqnum{10b}
\end{equation}

\begin{equation}
\Delta E_{3c}=\frac{3}{4}\lambda ^{\prime }+\frac{9}{32} \lambda
\eqnum{10c}
\end{equation}

\begin{equation}
\Delta E_{3d}=\Delta E_{3e}=\frac{\Delta
_{3b}}{2}=\frac{1}{2}\left[ -\frac{%
3}{4}\lambda ^{\prime }+\frac{9}{32}\lambda \right]
\eqnum{10d-e}
\end{equation}

As we see from (10),
 the domain wall 3c is the most ``expensive'' one, it always has
 positive  creation energy. The situation with other three types of
 domain walls, 3b, 3d, and 3e in the notation of \cite{Hill}, is
 different:
 they cost us some energy for small $(\lambda ^{\prime }/\lambda )$,
 but they can be created spontaneously ($\Delta E<0$) if diagonal
 coupling $\lambda ^{\prime }$ is sufficiently strong.
 As we see from (10),
 the critical value for this coincides with the critical value
 $(\lambda ^{\prime }/\lambda )=3/8$ (9) obtained above, which
 confirms the consistency of our treatment.

 An important conclusion is that, at least in our model, in the range of
 existence of the CE structure,
 $(\lambda ^{\prime }/\lambda )<(\lambda ^{\prime }/\lambda )_{c}=3/8$,
 the domain walls of the type 3d and 3e are easiest to form, and they
 have the
 same creation energy. Thus, it will be presumably orbital twins 3d and
 3e which will appear in real systems. The driving force for their
 formation is apparently the same as in ordinary martensitic transitions
 \cite{Khach}, i.e. they are elastic domains caused by strain in
 crystals (this is suggested also by A. Millis, as cited in
 \cite{Hill}).
 different from what was usually assumed: instead of the

\section{Stripes and bistripes in overdoped manganites}
 Finally, we turn to the most controversial problem -- that of the
 type and the origin of superstructures in overdoped manganites, e.g.\
 stripes and bistripes in La$_{1-x}$Ca$_{x}$MnO$_{3}$ at $x=2/3$; 3/4
 \cite{Radaelli,Raveau,ChenCheong}. First of all, there exists an
 experimental controversy: different experiments give for the same system
 and even for the same samples different results. Thus, for $x=2/3$ the
 high-resolution electron microscopy gave the results taken as a
 signature of paired stripes or bistripes \cite{ChenCheong}, whereas the
 neutron scattering results were interpreted in terms of single stripes
 (called ``Wigner crystal'' in \cite{Radaelli}).

%%%%%%%%%%%*****************FIG.6*******************%
\begin{figure}[tbp] \epsfxsize= .95\hsize  \vskip 1.0\baselineskip
\centerline{\epsffile{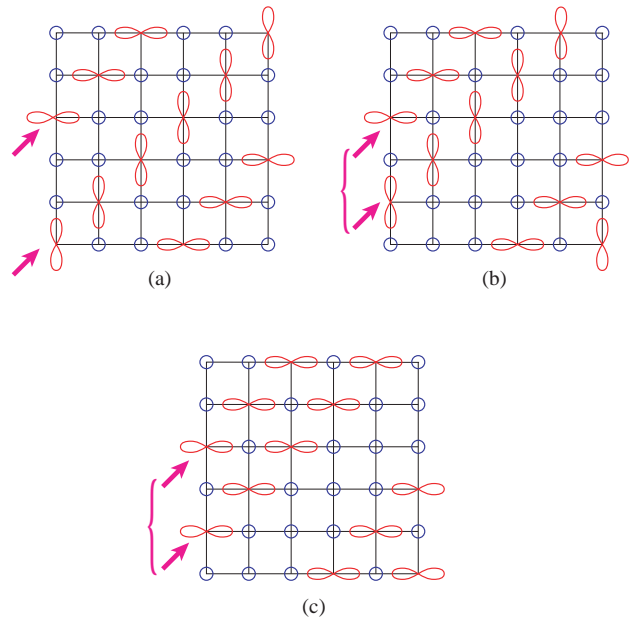}}
  \vskip1\baselineskip plus1\baselineskip
\caption{Possible stripe structures in the basal plane of
 La$_{1-x}$Ca$_{x}$MnO$_{^{3}}$ at $x=2/3$: (a) single stripes;
 (b) bistripes with different orbitals; (c) bistripes with
 similar orbitals.}
\label{stripes}
\end{figure}
%%%%%%%%%%%%%%%%%%%%%%%%%%%%%%%%%%%%%%%%%%%%%%%%%

 These two situations,
 with the corresponding orbital ordering, are illustrated in Fig.\
 6a,~b for the case $x=2/3$. The orbitals at each stripe are
 parallel, e.g.\ one stripe is formed by $x^{2}$ orbitals and another
 stripe by $y^{2}$ orbitals.
 In the paired stripe picture (Fig.~6b), these two diagonal stripes,
 with $x^{2}$ and $y^{2}$ orbitals, come close together being
 separated by one Mn$^{4+}$ diagonal row.

Again, we have here several questions: why stripes here are better
 than any other type of charge ordering? (e.g., for $x=3/4$, one can
 arrange the real ``Wigner crystal'' structure with the Mn$^{3+}$ sites
 forming a face-centered 2D lattice); and why do stripes have a particular
 orbital structure? And the third, most burning question is which state,
 single or paired stripes, is better?

One can try to approach these problems using our model with
 the interactions of Table 1, adding to it a restriction forbidding
 the formation of dense Mn$^{3+}$ clusters. Similar to Section VI,
 we calculate now the energies of the single stripes of Fig.~6a,
 paired stripes of Fig.~6b, and also the possible competing type of
 bistripes shown in Fig.~6c, with all orbitals of the same type,
 e.g.\ $x^2$. Using the interaction parameters of Table 1 and keeping
 again the interaction of the first (diagonal) and of the second
 ``direct'' neighbours at a distance of two lattice constants in $x$
 and $y$ directions, we obtain:
\begin{equation}
E_{6a}=-\frac{\lambda ^{\prime }}{3}  \eqnum{11}
\end{equation}
\begin{equation}
E_{6b}=-\frac{\lambda ^{\prime }}{3}+\frac{1}{2}\frac{5\lambda
^{\prime }}{12%
}+\frac{1}{8}\left( -\frac{\lambda }{2}\right) =-\frac{\lambda
^{\prime }}{8}%
-\frac{\lambda }{16},  \eqnum{12}
\end{equation}
\begin{equation}
E_{6c}=\frac{3}{2}\left(-\frac{\lambda ^{\prime }}{3}\right)+%
\frac{1}{2}\left( \frac{\lambda }{8}+%
\frac{1}{8}\frac{\lambda }{4}\right) =-\frac{\lambda ^{\prime }}{2}%
+\frac{5\lambda }{64}.  \eqnum{13}
\end{equation}
Expression (11) is quite clear. In (12), the first term describes
 the diagonal interaction along the stripe, the second  -- diagonal
 interactions between stripes in the pair (each such bond belongs to
 two sites, therefore we have factor 1/2 in it), and the third term
 is the next-nearest-neighbour interaction of $x^{2}$ orbital of one
 stripe with the $y^{2}$ orbitals of its ``mate''. Similarly, one
 can understand the meaning of different terms in (13).

 Comparing the energies of these three states, one obtains that there
 exist two critical values of parameter
 $\nu =\lambda ^{\prime }/\lambda $

\begin{equation}
\nu _{{\normalsize cr}}^{(ab)}=3/10=0.3,  \eqnum{14}
\end{equation}
\begin{equation}
\nu _{{\normalsize cr}}^{(ac)}=15/32\simeq0.47.  \eqnum{15}
\end{equation}
For $\lambda ^{\prime }/\lambda <\nu _{cr}^{(ab)}=0.3$, the paired
 stripes of Fig.~6b are stable. For $0.3=\nu _{cr}^{(ab)} <
\lambda ^{\prime }/\lambda <\nu _{cr}^{(ac)}=0.47$, the single-stripe
 phase of Fig.~6a would have the lowest energy among the three states,
 which we compare. And finally, for $ \lambda ^{\prime }/\lambda <
 \nu _{cr}^{(ac)}=0.47$, the paired stripes of Fig.~6c would have the
 lowest energy. We again discard here the possibility of phase
 separation, where all electrons are assembled in the bulk CE-like
 phase of Fig.~5a, which can arise at small $\lambda ^{\prime  }/\lambda $,
 and into the phase of the type shown in Fig.~5b for
 large $\lambda ^{\prime }/\lambda $, assuming that such phase
 separation into electron-enriched and electron-depleted regions
 would be prevented, e.g.\ by the long-range Coulomb interaction.

Comparing these results with those of Section VIa (half-doped systems),
 we can see that for realistic values of $\lambda ^{\prime }/\lambda  <0.375$
 (9), which are needed to stabilize the experimentally observed
 CE phase, we may have either the paired stripes \cite{ChenCheong}
 (Fig.~6b), or single stripe phase of Fig.~6a, see phase diagram in
 Fig.~7. Unfortunately,
 within this treatment we cannot make a unique choice between these two
 possibilities. Nevertheless, we can say that at least the sometimes used
 arguments \cite{ChenCheong}, which ascribe the pairing of stripes
 to strain interactions, are not very convincing: we see
 that the same strain interactions for different ranges of parameters
 may rather favour single stripes. Apparently, a more detailed treatment,
 including also long-range interactions, is needed to answer theoretically
 the question, which state, single or paired stripes, would be preferred for
 real systems (although it is, of course, primarily an experimental
problem).

%%%%%%%%%%%%%%%%%%%%%%%%%% Fig. 7 %%%%%%%%%%%%%%%%%%%%%%%%%%%%%%%%%%%%%
\begin{figure}[tbp]
\epsfxsize= .95\hsize  \vskip 1.0\baselineskip
\vbox{\centerline{\epsffile{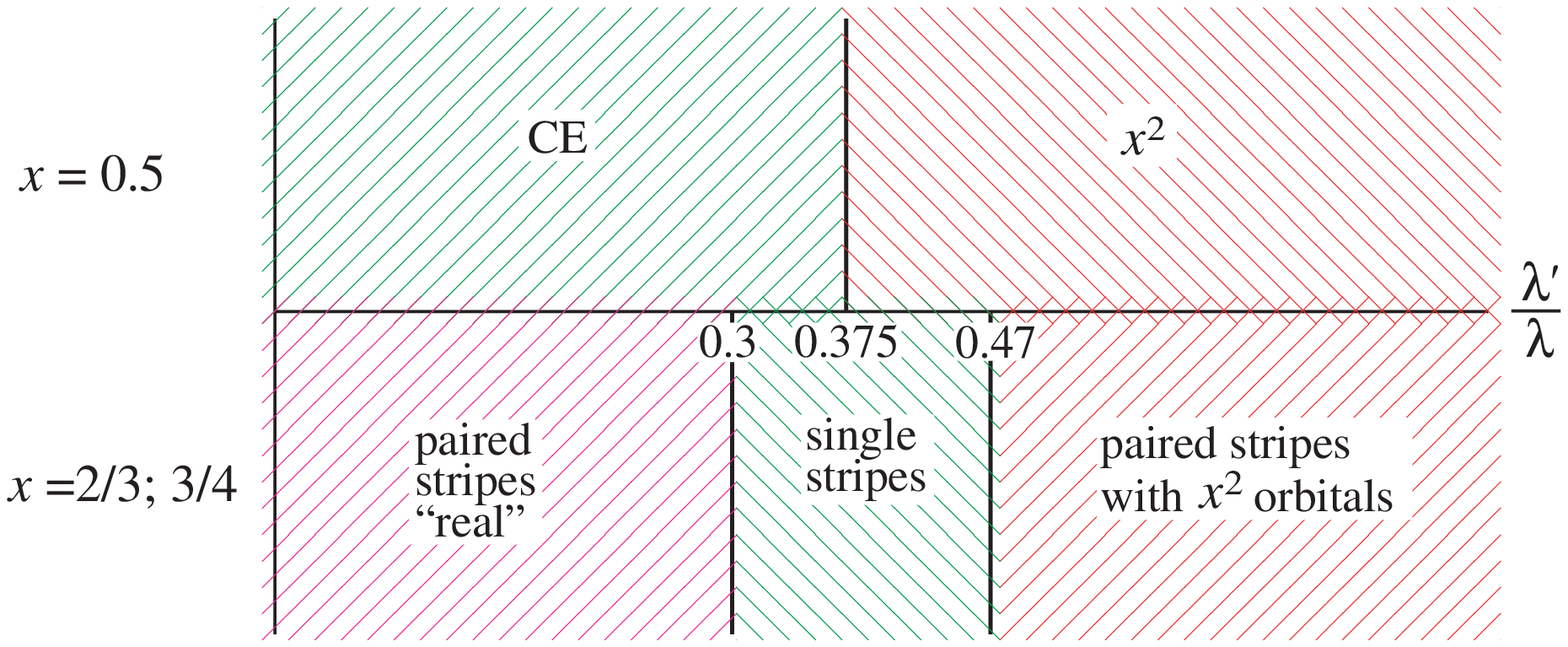}}
  \vskip1\baselineskip plus1\baselineskip
\caption{Regions of stability of different phases for $x=0.5$ (upper
 half-plane) and for overdoped manganites ($x=2/3$; 3/4, lower
 half-plane). Note that we do not consider fully phase-separated
 states.}}
\label{phdiag}
\end{figure}

As an indirect argument we may only use the results of the study of
 domain walls at the end of Section VI. As shown there, the energy
 cost of domain wall formation goes to zero at the critical point
 $(\lambda ^{\prime }/\lambda )=0.375$ (9).
 As follows from the experimental data \cite{Hill}, orbital domains
 are formed very easily in half-doped manganites. It may be a signature
 that real systems are not far from this critical point. If true,
 this would mean that, according to the results (12) and (13) and
 to Fig.~7, single stripes may be preferable, although the energy
 of paired stripes may be not very far from it, so that they may become
 stable e.g. at the surface of the crystal. This could resolve the
 experimental controversy between the results of \cite{Radaelli}
 and~\cite{ChenCheong}.

\section{Modeling of orbital ordering by interacting quadrupoles}

As is well known \cite{KugKhUFN}, the orbital ordering is closely
 related to the ordering of electric quadrupoles: the asymmetric
 distribution of charge density at a transition metal ion with
 a particular occupation of one of degenerate orbitals corresponds
 to a quadrupolar charge distribution, and, moreover, the interaction
 of orbitals via lattice distortions is very similar to the conventional
 quadrupole-quadrupole (QQ) interaction. It is instructive to see whether
 one can model the elastic interactions listed in Table 1 by this QQ
 interaction. Of course, in real transition metal compounds this
 direct QQ interaction is much weaker than that via phonons (it may
 be relevant though for rare earths compounds, where one indeed
 often uses the terminology ``quadrupolar ordering'' instead of our
 ``orbital ordering`` and ``cooperative Jahn-Teller effect'').
 The distance dependence is also different: the QQ interactions
 decay as $1/R^{5}$, whereas the interactions via phonons -- only
 as $1/R^{3}$.  Nevertheless, the operator structure of the
 corresponding terms in the Hamiltonian is very similar, and one
 may expect that at least the signs, and probably the ratios of
 corresponding coupling constants would be similar for both
 these mechanisms.

The general form of interaction between two quad\-ru\-poles
 Q$_{\alpha \beta }$ and Q$_{\gamma \delta }$, located in positions
 with radius vector {\bf R} connecting them, can be obtained from the
 general treatment \cite{QQ}:
\begin{eqnarray}
U_{QQ'%
} &=&\frac{5}{12R^{5}}\Biggl[7Q_{\alpha \beta }Q'_{\gamma \delta }
n_{\alpha }n_{\beta }n_{\gamma }n_{\delta }-  \nonumber \\
&&4Q_{\alpha \beta }Q'_{\alpha \gamma }
n_{\beta }n_{\gamma }+\frac{2}{5}Q_{\alpha \beta }Q'_{\alpha \beta }
\Biggr]dV  \eqnum{16}
\end{eqnarray}
where $n_{\alpha}$ {\it etc.}\ are directional cosines of
 radius-vector~{\bf R}.

 Using this general expression for axially symmetric quadrupoles
 ($Q_{zz} = Q$, $Q_{xx} = Q_{yy} = -Q/2$), one can easily calculate
 the quadrupolar interaction constants for the cases 1) -- 8) of
 Table~1. They turn out to be
  \begin{eqnarray}
V_{1)} = \Lambda; V_{2)}= - \frac{\Lambda}{2};\quad V_{3)}=
 \frac{\Lambda}{8};\quad V_{4)}= \frac{3 \Lambda}{8}; \nonumber \\
V_{5)} = - \frac{1}{(\sqrt{2})^{5}} \frac{13 \Lambda}{40};\quad
V_{6)}= \frac{1}{(\sqrt{2})^{5}} \frac{49 \Lambda}{96}; \nonumber \\
 V_{7)}= \frac{1}{(\sqrt{2})^{5}} \frac{3 \Lambda}{8};\quad
 V_{8)}= - \frac{1}{(\sqrt{2})^{5}} \frac{3 \Lambda}{8} ,
 \eqnum{17}
\end{eqnarray}
 where $\Lambda =3Q^{2}/2R^{5}$. In calculating the values of
 $V_{1)} - V_{8)}$ in (17), we separated out the factor
 $1/(\sqrt{2})^{5}$, since the distance for diagonal pairs is larger
 by a factor of $\sqrt{2}$ than that for the ``direct'' nearest
 neighbours.

 One can see that indeed the signs of the QQ interactions in (17)
 coincide with those of elastic interactions (see Table 1); this
 gives an extra support to these values used in the main part of
 our paper. Moreover, even the ratios of the interaction constants
 are rather close, at least for cases 1) -- 4). In principle, one
 can try to find out the type of the ordered ground state for the
 QQ case, using general expression (16) and taking into account
 the longer-range interactions and interactions for all possible
 directions. This task would require rather extensive numerical
 calculations (or some clever summation of the Madelung type);
 we do not do this here. One can only formulate some interesting
 problems, which should be addressed in this field.

(1) Would there be indeed a long-range ordered state in this system?
 We remind that for a similar problem with the dipole, not quadrupole,
 interactions it was proven \cite{GlassLawson} using the approach
 similar to that of Mermin-Wagner \cite{Mermin} that there is no
 long-range ordering for the dipole-dipole system in the case of
 3D cubic lattice. Such possibility is not excluded also for the
 QQ interaction (for systems with quadrupoles at each lattice site).

(2) Another curious problem is what would be the situation with
 the mobile quadrupoles if their concentration is small (their
 number is less than the number of lattice sites). Such
 quadrupoles definitely attract each other at certain directions
 and for certain relative orientations of their main axes.
 Would they form structures like stripes, sheets, or some more
 complicated superstructures, and which in particular? Note that
 this possibility is rather similar to the one suggested in
 \cite{KugKhEPL} for elastic interactions of spherical ``impurities''
 in anisotropic media. Here, however, one considers the
 interaction of anisotropic centers in the isotropic space.
 Nevertheless, the outcome may be similar (actually the consideration
 of Section IV have much in common with this problem; but for the
 QQ case one can go beyond approximation of the nearest-neighbor
 interactions made above and consider the problem in its full
 complexity). Note also that similar question exists also for
 mobile dipoles.

These situations have not only an academic interest: one can
 imagine that similar superstructures of various kinds can arise
 due to these interactions, e.g.\ in the solutions of anisotropic
 molecules in certain solvents, etc.

The questions discussed in this Section have only indirect relation
 to the main problem considered in the present paper, but still we see
 much in common between these situations, at least in principle.

\section{Conclusions}

In conclusion, we considered in this paper the possibility of
 formation of different superstructures in insulating systems with
 Jahn-Teller ions like some cuprates, chromites, and especially
 manganites at different doping levels. We saw that even keeping
 only the interactions between a few nearest neighbors, we can
 successfully describe the formation of different orbital
 superstructures. In particular, using only the elastic interactions
 one can immediately obtain the observed ordering pattern of undoped
 manganites. By assuming the checkerboard charge ordering, we can easily
 explain by this mechanism the type of orbital ordering (CE structure)
 observed in half-doped manganites ($x=0.5$). Using the same physical
 arguments, we suggested that the superstructures observed in the
 low-doped manganites (ferromagnetic insulating phases at $x=0.1-0.3$)
 involve the segregation of doped holes in consecutive planes,
 this structure being
 favored by the special stability of the orbital ordering of the
 ``undoped LaMnO$_{3}$-type'' -- the ordering shown in Fig.~2.
 For the overdoped manganites like La$_{1-x}$Ca$_{x}$MnO$_{3}$,
 $x > 0.5$, we argued that the stability of stripes or bistripes observed
 in this doping range (especially at $x=2/3$; 3/4) is also determined
 by the elastic forces, although we were not able to determine uniquely
 which of these alternatives, single or paired stripes, is more
 favorable. In our approach, the result depends on the ratio of two
 constants related to the noncentral and central forces
 respectively, which we do not know {\em a priori}. But in any case,
 the elastic forces largely determine the very tendency to the stripe
 formation in insulating overdoped manganites, as well as their orbital
 structure.

As mentioned in the Introduction, there exist also other possible
 mechanisms of orbital ordering, in particular, the superexchange
 interaction \cite{KugKhUFN,KugKhJETP}. Typically both these
 mechanisms, the exchange interaction and the interaction via lattice,
 lead to the same orbital structures in cases when we have
 Jahn-Teller ions at each site. It is quite difficult to find the
 cases, in which the outcomes of these two models would be qualitatively
 different \cite{KugKhFTT}. In some sense, it is gratifying that very
 simple considerations presented above immediately give the correct
 orbital structures for the cases like LaMnO$_{3}$ or K$_{2}$CuF$_{4}$,
 whereas rather hard work is required to get these structures with
 the electronic (exchange) mechanism. Thus, apparently both these mechanisms
 are operational in real systems, and the relative importance of each is
 still an open problem, at least for the cases of dense systems (see
 also \cite{Anisimov} in this context).

But when we turn to the dilute systems, like doped manganites,
 especially in the overdoped regime $x>0.5$, then apparently the
 elastic interactions play the most important role, first of all
 providing the essential mechanism of stripe formation, and also
 determining their orbital structure: one can expect that the exchange
 interactions, being essentially short-range, are less efficient in these
 cases. It is mainly in these systems, where we believe that our approach
 may be most fruitful.

\section*{Acknowledgments}

The work was supported by INTAS (grants 97-0963 and 97-11954),
 the Russian Foundation for Basic Research (grant 00-15-96570),
 the Netherlands Foundation for the Fundamental Research of Matter (FOM),
 and the Netherlands Organization for the Advancement of Pure
 Research (NWO).

%\section{REFERENCES}


\begin{references}
\bibitem{Goodenough}   J.~B.~Goodenough, {\em Magnetism and the
 Chemical Bond} (Interscience, New York, London, 1963).
\bibitem{KaplVekh}   M.~D.~Kaplan and B.~G.~Vekhter, {\em Cooperative
 Phenomena in Jahn-Teller Crystals} (Plenum, New York, 1995).
\bibitem{KugKhUFN}  K.~I.~Kugel and D.~I.~Khomskii, Sov.
 Phys. -- Uspekhi {\bf 25}, 231 (1982).
\bibitem{KhSawats}  D.~I.~Khomskii and G.~Sawatzky, Solid State. Comm.
 {\bf 102}, 87 (1997).
\bibitem{KhSpinEl}  D.~I.~Khomskii, chapter in {\em Spin Electronics},
 eds. M.~Ziese and M.~J.~Thornton, (Springer, Berlin, 2001), p. 89.
\bibitem{Radaelli}  P.~G.~Radaelli, D.~E.~Cox, L.~Capogna, S.-W.~Cheong,
 and M.~Marezio, Phys. Rev. B {\bf 59}, 14440 (1999).
\bibitem{Raveau}B.~Raveau, M.~Hervieu, A.~Maignan, and C.~Martin,
J. Mater. Chem. {\bf 11}, 29 (2001).
\bibitem{ChenCheong} S.~Mori, C.~H.~Chen, and S.-W.~Cheong, Nature
 (London) {\bf 392}, 473 (1998).
\bibitem{Egami}  T.~Egami and D.~Louca, Phys. Rev. B {\bf 59},
 6193 (1999).
\bibitem{vdBrinkKh}  D.~I.~Khomskii, cond-mat/0004034 (2000); 
 J.~van~den Brink and D.~I.~Khomskii, Phys. Rev. B {\bf 63},
 140416 (2001).
\bibitem{Englman}  R.~Englman, {\em The Jahn-Teller Effect in
 Molecules and Crystals} (Wiley-Interscience, New York, 1972).
\bibitem{EnglHalp}  R.~Englman and B.~Halperin, Phys. Rev. B {\bf 2},
 75 (1970); B. Halperin and R. Englman, Phys. Rev. B {\bf 3},
 1698 (1971)
\bibitem{KugKhJETP}  K.~I.~Kugel and D.~I.~Khomskii, JETP Lett.
 {\bf 15}, 446 (1972); K.~I.~Kugel and D.~I.~Khomskii, Sov. Phys.
 -- JETP {\bf 37}, 725 (1973).
\bibitem{KugKhEPL}  D.~I.~Khomskii and K.~I.~Kugel, Europhys. Lett.
 {\bf 55}, 208 (2001).
\bibitem{Eshelby}  J.~D.~Eshelby, {\em The continuum theory of lattice
 defects} in {\em Solid State Physics}, eds. F.~Seitz and D.~Turnbull,
 {\bf 3}, 79 (Academic Press, New York, 1956).
\bibitem{Eremin}  M.~V.~Eremin, A.~Yu.~Zavidonov, and B.~I.~Kochelaev,
 Zh. Eksp. Teor. Fiz. {\bf 90}, 537 (1986) [Sov. Phys. -- JETP {\bf 63},
 312 (1986)].
\bibitem{Fishman}  M.~A.~Ivanov, V.~Ya.~Mitrofanov, and A.~Ya.~Fishman,
 Fiz. Tv. Tela (Sov.Phys. -- Solid State) {\bf 20}, 3023 (1978).
\bibitem{Kanamori}  J.~Kanamori, J. Appl. Phys. (Suppl.) {\bf 31},
 14S (1960).
\bibitem{KhvdBr-anh}  D.~I.~Khomskii and J.~van~den~Brink, Phys. Rev.
 Lett. {\bf 85}, 3329 (2001).
\bibitem{Yto}  Y.~Yto and J.~Akimitsu, J. Phys. Soc. Japan {\bf 40},
 1333 (1976).
\bibitem{KhKugSSC}  D.~I.~Khomskii and K.~I.~Kugel, Solid State Comm.
 {\bf 13}, 763 (1973).
\bibitem{RbCuCl3}  M.~T.~Hutchins, M.~J.~Fair, P.~Day, and
 P.~J.~Walker, J. Phys C {\bf 9}, L55 (1976).
\bibitem{z^2order}  T.~G.~Perring, D.~T.~Androja, G.~Chaboussant,
 G.~Aeppli, T.~Kimura, and Y.~Tokura, Phys. Rev. Lett. {\bf 82},
 217201 (2001).
\bibitem{Novak}  P.~Novak, J. Phys. Chem. Solids {\bf 30},
 2357 (1969); {\bf 31}, 125 (1970).
\bibitem{Mizokawa}  T.~Mizokawa, D.~I.~Khomskii, and G.~A.~Sawatzky,
 Phys. Rev. B {\bf 60}, 7309 (1999).
\bibitem{JapLaSr(1/8)}  H.~Kawano, R.~Kajumoto, M.~Kubota,
 and H.~Yoshizawa, Phys. Rev. B {\bf 53}, 2202 (1996);
{\bf 53}, R14709 (1996)
\bibitem{BuechLaSr(1/8)}  S.~Uhlenbruck, R.~Teipen, R.~Klingeler,
 B.~B\"{u}chner, O.~Friedt, M.~H\"{u}cker, H. Kierspel,
 T. Niem\"{o}ller, L.~Pinsard, A.~Revcolevschi, and R.~Gross,
 Phys. Rev. Lett. {\bf 82}, 185 (1999)
\bibitem{Yamada}  Y.~Yamada, O.~Hino, S.~Nohdo, R.~Kanao, T.~Inami,
 and S.~Katano, Phys. Rev. Lett. {\bf 77}, 904 (1996).
\bibitem{Mizokawa1}  T.~Mizokawa, D.~I.~Khomskii, and G.~A.~Sawatzky,
 Phys. Rev. B {\bf 61}, R3776 (2000).
\bibitem{Kilian}  R.~Kilian and G.~Khaliullin, Phys. Rev. B {\bf 60},
 13458 (1999).
\bibitem{Mizokawa2}  T.~Mizokawa, D.~I.~Khomskii, and G.~A.~Sawatzky,
 Phys. Rev. B {\bf 63}, 024403 (2001).
\bibitem{AhnMillis}  K.~H.~Ahn and A.~J.~Millis, Phys. Rev. B
 {\bf 58}, 3697 (1998).
\bibitem{Tomioka}  Y.~Tomioka, A.~Asamitsu, H.~Kawahara, Y.~Moritomo,
 and Y.~Tokura, Phys. Rev. B {\bf 53}, 1689 (1996).
\bibitem{orbdomains}P.~G.~Radaelli, D.~E.~Cox, M.~Marezio,
 and S.-W.~Cheong, Phys. Rev. B {\bf 55}, 3015 (1997).
\bibitem{Zimmermann}  M.~v.~Zimmermann, C.~S.~Nelson, J.~P. Hill,
 D.~Gibbs, M.~Blume, D.~Casa, B.~Keimer, Y.~Murakami, C.-C.~Kao,
 C.~Venkataraman, T.~Gog, Y.~Tomioka, and Y.~Tokura,
 Phys. Rev. B {\bf 64}, 195133 (2001).
\bibitem{Hill}  J.~P.~Hill, C.~S.~Nelson, M.~v.~Zimmermann,
 Y.-J. Kim, D.~Gibbs, D.~Casa, B.~Keimer, Y.~Murakami,
 C.~Venkataraman, T.~Gog, Y.~Tomioka, Y.~Tokura, V.~Kiryukhin,
 T.~Y.~Koo, and S.-W.~Cheong, Appl. Phys. A {\bf 73}, 723 (2001);
 cond-mat/0105064.
\bibitem{vdBrKhalKh}  J.~van~den~Brink, G.~Khaliullin, and
 D.~I.~Khomskii, Phys. Rev. Lett. {\bf 83}, 5118 (1999).
\bibitem{Wollan}  E.~O.~Wollan and W.~C.~Koehler, Phys. Rev.
 {\bf 100}, 545 (1955).
\bibitem{Jirak}  Z.~Jir\'ak, S.~Krupi\v cka, Z.~\v Sim\v sa,
 M.~Dlouh\' a, and S.~Vratislav, J. Magn. Magn. Mater. {\bf 53},
 153 (1985).
\bibitem{Khach}  A.~G.~Khachaturyan, {\em Theory of Phase
 Transformations and the Structure of Solid Solutions}
 (Nauka, Moscow, 1974).
\bibitem{Kagan}  M.~Yu.~Kagan, K.~I.~Kugel, and D.~I.~Khomskii,
 Zh. Exp. Teor. Fiz. {\bf 120}, 470 (2001) [JETP {\bf 93},
 415 (2001)]; cond-mat/0001245.
\bibitem{QQ}  L.~D.~Landau and E.~M.~Lifshitz, {\em The Classical
 Theory of Fields} (Butterworth-Heinemann, Oxford, 1997).
\bibitem{GlassLawson}  S.~J.~Glass and J.~O.~Lawson, Phys. Lett. A
 {\bf 46}, 234 (1973).
\bibitem{Mermin}  N.~Mermin and H.~Wagner, Phys. Rev. Lett. {\bf 17}
 1133 (1966).
 \bibitem{KugKhFTT}  K.~I.~Kugel and D.~I.~Khomskii, Fiz.
 Tv. Tela (Sov. Phys. -- Solid State) {\bf 15}, 2230 (1973).
\bibitem{Anisimov}  A.~I.~Liechtenstein, V.~I.~Anisimov, and J.~Zaanen,
 Phys. Rev. B {\bf 52}, R5467 (1995).
\newpage
\end{references}
\end{document}